\documentclass[a4paper]{article}[12pt]
\usepackage[english]{babel}
\usepackage{float}
\usepackage{amsmath,amssymb,color}
\usepackage{amsfonts}
\usepackage{amsthm}
\usepackage{caption}
\usepackage{subcaption}
\usepackage{graphicx}
\usepackage{natbib}
\usepackage{hyperref}
\usepackage[colorinlistoftodos]{todonotes}
\usepackage[top=1.2in, bottom=1.2in, left=1in, right=1in]{geometry}
\normalsize

\newcommand{\ttt}{\boldsymbol \theta}

\DeclareMathOperator*{\argmax}{arg\,max}

\title{Estimation Methods for Item Factor Analysis: An Overview}

\author{Yunxiao Chen and Siliang Zhang\\
London School of Economics and Political Science}

\date{}

\begin{document}
\maketitle

\begin{abstract}
Item factor analysis (IFA) refers to the factor models and statistical inference procedures for analyzing multivariate categorical data.
IFA techniques are commonly used in social and behavioral sciences for analyzing item-level response data. Such models summarize and interpret the dependence structure among a set of categorical variables by a small number of latent factors. In this chapter, we review the IFA modeling technique and commonly used IFA models. Then we discuss estimation methods for IFA models and their computation,
with a focus on the situation where the sample size, the number of items, and the number of factors are all large. Existing statistical softwares for IFA are surveyed. This chapter is concluded with suggestions for practical applications of IFA methods and discussions of future directions.

\end{abstract}	


\section{Introduction} 
\label{sec:outline_of_this_chapter}

Item-level response data are commonly encountered in social and behavioral sciences, including education, psychology, psychiatry, marketing, and political science.  Such data are typically binary (e.g., disagree/agree) or ordinal (e.g., strongly disagree, disagree,
neither, agree, and strongly agree), due to the way questionnaire items are designed. Due to the categorical nature of data, the traditional linear factor models are no longer suitable and item factor analysis models have been developed.

IFA models combine the idea of common factor analysis and the generalized linear modeling techniques for categorical data.
The introduction of common factors allows IFA to characterize the joint distribution of a large set of observed variables by a smaller set of latent factors, leading to a reduction in data dimensionality.
The latent factors are often interpreted substantively as common causal factors, such as personality factors, general intelligence, mental health factors, political standings, etc. 
The IFA models further provide a characterization of the relationship between the common factors and the observed variables, through
the so-called item response functions which take a generalized linear model (GLM) regression form, where the response variables in the GLM regression are the observed variables (i.e., item responses) and the independent variables are the common factors.
Thanks to its good interpretation, IFA models are widely used for generating and testing substantive theory.

The last decade has observed the need of solving large-scale IFA problems, where the sample size, the number of items, and the number of latent factors can all be large. For example, in psychiatry, there has been a need to better understand and classify mental disorders \citep[e.g.,][]{kotov2017hierarchical}. This implies the need of applying IFA to many mental health symptoms from a large number of respondents, for which a large number of factors may be needed. In modern psychology, an important problem is to better characterize personality traits, for example by collecting large-scale data from the internet \citep[e.g.,][]{skitka2006internet,revelle2016web}. It means fitting IFA models to data with thousands of items and hundreds of thousands of respondents. In marketing, there are also needs to better analyze customers' preference choices by the IFA of large-scale e-commerce data. In this chapter, we will focus on reviewing estimation methods that are tailored to such large-scale IFA problems.


The rest of this chapter is organized as follows. In Section 2, we introduce the statistical frameworks of IFA. In Section 3, methods and algorithms for the estimation of IFA models are reviewed, followed by a survey of available computer softwares/packages in Section 4. We conclude this chapter with suggestions for the practical applications of IFA methods and discussions of future directions.


\section{IFA Models}

\subsection{Modeling Framework}\label{subsec:model}
Consider $N$ individuals answering $J$ items, with $Y_{ij}$ being the response from person $i$ to item $j$ and $Y = (Y_{ij})_{N\times J}$ being the data matrix. The data entry $Y_{ij}$ is typically categorical, due to the nature of item-level response data. In particular, $Y_{ij}$ takes values $0$ or $1$ for binary items, and $Y_{ij}$ takes value 0, 1, ..., $t_j$, for some $t_j \geq 2$ when the item is ordinal.
An IFA model imposes a joint distribution on the data matrix $Y = (Y_{ij})_{N\times J}$.
A typical IFA model makes the following three assumptions.
\begin{itemize}
  \item[A1.] Each individual is assumed to be represented by a latent trait vector $\ttt_i = (\theta_{i1}, ..., \theta_{iK})^\top$, for some $K \geq 1$. The distribution of person $i$'s responses $\mathbf Y_i = (Y_{i1}, ..., Y_{iJ})^\top$ depends only on $\ttt_i$ but not other $\ttt_{i'}$s, for $i'\neq i$.
      The dimension $K$ of the latent vector is typically chosen much smaller than the number of items $J$, so that a reduction in the data dimensionality is achieved. Depending on the types of applications, the value of $K$ is either determined by substantive theory (typically in confirmatory analysis) or to be estimated from data (typically in exploratory analysis). IFA models with $K = 1$ are often known as unidimensional models and those with $K > 1$ are known as multidimensional models.
  \item[A2.] Most IFA models assume local independence. That is, $Y_{i1},\ldots, Y_{iJ}$, are assumed to be conditionally independent, given the latent vectors $\ttt_i,$ for $i = 1, ..., N$.
  \item[A3.] Making use of the previous two assumptions, an IFA model completes the specification of the joint distribution of $Y$ by specifying the conditional distribution of each $Y_{ij}$ given $\ttt_i$. A parametric model is typically assumed, 
      $$P(Y_{ij} = t\vert \ttt_i; \boldsymbol\beta_j) = f_j(t\vert \ttt_i; \boldsymbol\beta_j),$$
      where $t\in \{0, 1\}$ for binary items
      and $t\in \{0, 1, ..., t_j\}$ for ordinal items, $\boldsymbol\beta_j$
      is a generic notation for the item-specific parameters, and $f_j$ is known as the item response function. Specific item response functions $f_j$ will be discussed in Section~\ref{subsec:example}.
\end{itemize}

These three assumptions lead to a joint distribution of the data matrix $Y$, given the person-specific latent factors $\ttt_i$ and the item specific parameters $\boldsymbol\beta_j$. 
This leads to the joint likelihood function given observed data $y_{ij}, i = 1, ..., N, j = 1, ..., J$,
\begin{equation}\label{eq:jl}
\begin{aligned}
&L_{\text{J}}(\ttt_i, \boldsymbol\beta_j, i = 1, \cdots, N, j = 1, \cdots, J)\\
=& \prod_{i=1}^N \prod_{j=1}^J f_j(y_{ij}\vert \ttt_i; \boldsymbol\beta_j).
\end{aligned}
\end{equation}
In this joint likelihood function, both the person-specific latent factors and the item-specific parameters are treated as unknown fixed parameters.  Alternatively, the latent factors are often treated as random effects instead of fixed parameters, under the following assumption.
\begin{itemize}
  \item[A4.] The latent vectors $\ttt_i$ are independent and identically distributed, following a cumulative distribution function $F$.
  A parametric form is typically assumed for $F$, where we use $\boldsymbol \gamma$ to denote the parameters and use $F(\cdot \vert \boldsymbol \gamma)$ to denote the parameterized cumulative distribution.
  In most IFA applications, $F$ is assumed to be multivariate normal.
\end{itemize}
This additional assumption, together with the previous assumptions, implies the marginal likelihood function
\begin{equation}\label{eq:ml}
\begin{aligned}
&L_{\text{M}}(\boldsymbol\gamma, \boldsymbol\beta_j, j = 1, \cdots, J)\\
=& \prod_{i=1}^N \int \left[\prod_{j=1}^J f_j(y_{ij}\vert \ttt; \boldsymbol\beta_j)\right]\phi(\ttt\vert \boldsymbol \gamma)d\ttt,
\end{aligned}
\end{equation}
where $\phi(\ttt\vert \boldsymbol \gamma)$ denotes the density function of $F$. Fundamentally, the
two different views of the latent factor come from different sampling foundations of the IFA models. In particular, the fixed effect view of the latent factors has a ``stochastic subject" interpretation and the random effect view has a
``random sampling" interpretation of the probability in IFA models. See \cite{holland1990sampling} for detailed discussions.
Technically, as will be discussed in Section~\ref{sec:estimation}, the estimations based on the two likelihood functions have different asymptotic behaviors.

Although our focus is on multivariate categorical data, the above modeling framework also includes the linear factor models as a special case. Specifically, a linear factor model takes the form of
\begin{equation}\label{eq:linear}
Y_{ij} = d_j + \mathbf a_j^\top \ttt_i + \epsilon_{ij},
\end{equation}
where $Y_{ij}$ is a continuous variable and $\epsilon_{ij}$ is a mean-zero independent error term with variance $\sigma_j^2$. 

\subsection{Examples of IFA Models}\label{subsec:example}

In what follows, we discuss some specific IFA models. We separate the discussion by the type of data.

\paragraph{Models for binary data.} When $Y_{ij}$ is binary, one only needs to specify $P(Y_{ij} = 1\vert \ttt_i; \boldsymbol\beta_j) = f(1\vert \ttt_i; \boldsymbol\beta_j)$, which can be viewed as the specification of a generalized linear model for binary response,
with $Y_{ij}$ being the response and $\ttt_i$ being the independent variables.
A commonly used parametrization is
\begin{equation}\label{eq:binary}
f(1\vert \ttt_i; \boldsymbol\beta_j) = G(d_j + \mathbf a_j^\top \ttt_i),
\end{equation}
where $G: \mathbb R \rightarrow (0, 1)$ is a pre-specified monotonically increasing function and item parameters $\boldsymbol \beta_j = (d_j, \mathbf a_j)$.
Function $G$ is often known as the inverse link function, using the terminology of generalized linear models. Commonly used choices of $G$ include the logistic and probit forms, for which
$G(x) = \exp(x)/(1+\exp(x))$ and $G(x) = (\int_{-\infty}^x \exp(-z^2/2) dz)/\sqrt{2\pi}$, respectively.
When viewing \eqref{eq:binary} as a generalized linear model that regresses $Y_{ij}$ on $\ttt_i$, $d_j$ can be viewed as the intercept parameter and $\mathbf a_j = (a_{j1}, ..., a_{jK})^\top$ are the slope parameters. The slope parameters are also known as the loading parameters in IFA.

When $K = 1$ and $G(x)$ takes the logistic form, then the model
$$f(1\vert \ttt_i; \boldsymbol\beta_j) = \frac{\exp(d_j + a_j\theta_i)}{1+\exp(d_j + a_j\theta_i)}$$
is known as the two-parameter logistic model (2PL; \citealp{birnbaum1968some}).  When $a_j$ is further restricted to be 1, then the model
$$f(1\vert \ttt_i; \boldsymbol\beta_j) = \frac{\exp(d_j + \theta_i)}{1+\exp(d_j + \theta_i)}$$
becomes a reparametrization of the famous Rasch model \citep{rasch1960probabilistic}.  Both models are widely used in the educational testing industry for the design, analysis, and scoring of tests.
When $K > 1$, both the logistic and probit versions of model~\eqref{eq:binary} are commonly used for multidimensional IFA analysis, with the logistic version typically known as the multidimensional two-parameter logistic model (M2PL; \citealp{reckase2009multidimensional}).
As the two link functions can approximate each other well \citep[see e.g.,][]{birnbaum1968some}, IFA result from the logistic model and that from the probit model are usually very similar. Therefore, in practice,  the choice of the link function is typically determined 
by 
computational consideration. Roughly speaking, the logistic form tends to be easier to handle computationally when treating the latent factors as fixed parameters, and the probit form has advantages under the random effect view. Further discussions will be provided in Section~\ref{sec:estimation}.

Equivalently, model~\eqref{eq:binary} can be obtained through the introduction of a latent response. That is, we define
a latent response
\begin{equation}\label{eq:uv}
Y_{ij}^* = d_j + \mathbf a_j^\top \ttt_i + \epsilon_{ij},
\end{equation}
where $\epsilon_{ij}$ is an independent error term with zero mean.
This equation takes the same form as \eqref{eq:linear} for linear factor model, except that the observed variables in \eqref{eq:linear} is replaced by the latent response in \eqref{eq:uv}.
The observed response is assumed to be a truncated version of the latent response, i.e.,
\begin{equation}\label{eq:uv2}
Y_{ij} = 1_{\{Y_{ij}^* \geq 0\}}.
\end{equation}
When $\epsilon_{ij}$ follows the standard normal and logistic distributions, respectively, \eqref{eq:uv} and \eqref{eq:uv2} together imply
the probit and logistic versions of \eqref{eq:binary}, respectively.
The latent response formulation brings computational convenience for the probit model, when $\ttt_i$s are viewed as random effects and follow a multivariate normal distribution. This computational advantage is brought by a data argumentation trick for Monte Carlo sampling; see Section~\ref{subsec:mml} for the details.

\paragraph{Models for ordinal data.} Ordinal responses are probably even more common than binary responses in practice due to the wide use of Likert-scale items in social and behavioral sciences. Models for ordinal data naturally generalize those for binary data by two different approaches. 
The first approach obtains $f_j(t\vert \ttt_i; \boldsymbol\beta_j)$ by specifying the cumulative probabilities of a response less than or equal to each threshold $t$. That is,
\begin{equation}\label{eq:ordinal}
P(Y_{ij} \leq  t \vert \ttt_i; \boldsymbol\beta_j)  = G(d_{jt} + \mathbf a_j^\top \ttt_i), ~t = 0, ..., t_{j-1},
\end{equation}
where, the same as in \eqref{eq:binary}, $G$ is a pre-specified monotonically increasing function, and $\boldsymbol\beta_j = (d_{j0}, ..., d_{j,t_j-1}, \mathbf a_j)$ are the item specific parameters. Recall that $Y_{ij}\in \{0,\ldots,t_j\}$, and thus $P(Y_{ij}\leq t_j\vert \ttt_i; \boldsymbol\beta_j)=1$. It is worth noting that the intercept parameter in \eqref{eq:ordinal} depends on the response category while the slope parameter does not. Similar as the binary case, the inverse link function $G$ is often chosen to take logistic or probit forms.

Due to the facts that $P(Y_{ij} \leq  t+1 \vert \ttt_i; \boldsymbol\beta_j) \geq P(Y_{ij} \leq  t \vert \ttt_i; \boldsymbol\beta_j)$ for each $t$ and that $G$ is monotonically increasing, we naturally have the constraints for the intercept parameters,
$$d_{j0}\leq d_{j1} \leq \cdots \leq d_{j,t_j-1}.$$
The cumulative probabilities \eqref{eq:ordinal} then imply the category-specific probabilities. That is,
$$f(t\vert \ttt_i; \boldsymbol\beta_j) = P(Y_{ij} \leq  t \vert \ttt_i; \boldsymbol\beta_j)  - P(Y_{ij} \leq  t-1 \vert \ttt_i; \boldsymbol\beta_j).$$
Depending on the dimension of the latent vector, the model \eqref{eq:ordinal} is referred to as the unidimensional and multidimensional graded response models \citep{Samejima1969estimation,muraki1995full}. The model~\eqref{eq:ordinal} has a similar latent response interpretation as the model for binary data. Therefore, it shares the same connection with linear factor models and has same computational advantage.

The second modeling approach is by specifying the conditional distributions based on the adjacent response categories. More precisely, the following conditional probabilities are specified
\begin{equation}\label{eq:ordinal2}
P(Y_{ij} =  t \vert \ttt_i,\boldsymbol\beta_j, Y_{ij} \in \{t-1, t\})  = G(d_{jt} + \mathbf a_j^\top \ttt_i), ~t = 1, ..., t_{j}.
\end{equation}
Again, $G$ is a pre-specified monotonically increasing function, and $\boldsymbol\beta_j = (d_{j1}, ..., d_{jt_j}, \mathbf a_j)$ are the item specific parameters. 
In this model, the inverse link $G$ is chosen to be the logistic form instead of the probit form.
As a result,
the conditional probability \eqref{eq:ordinal2} implies that
\begin{equation}\label{eq:odds}
\frac{f(t\vert \ttt_i; \boldsymbol\beta_j)}{f(t-1\vert \ttt_i; \boldsymbol\beta_j)} = \frac{G(d_{jt} + \mathbf a_j^\top \ttt_i)}{1-G(d_{jt} + \mathbf a_j^\top \ttt_i)} = \exp(d_{jt} + \mathbf a_j^\top \ttt_i).
\end{equation}
Note that the ratio of the probabilities in two adjacent categories does not have a simple form when using the probit link.
The ratio \eqref{eq:odds} further leads to
\begin{equation}\label{eq:pcm}
f(t\vert \ttt_i; \boldsymbol\beta_j) = \left\{\begin{array}{ll}
                                       {1}/\big({1 + \sum_{s=1}^{t_j} \exp(s\mathbf a_j^\top \ttt_i + \sum_{v=1}^s d_{jv})}\big), & \mbox{~if~} t = 0 \\
                                     {\exp(t\mathbf a_j^\top \ttt_i + \sum_{v=1}^t d_{jv})}/\big({1 + \sum_{s=1}^{t_j} \exp(s\mathbf a_j^\top \ttt_i + \sum_{v=1}^s d_{jv})}\big), & \mbox{~if~} t = 1, ...,t_j.
                                       \end{array}\right.
\end{equation}
This model is known as the generalized partial credit model \citep{muraki1992generalized,yao2006multidimensional}.

The generalized partial credit model and the graded response models tend to have similar fit on empirical data. Computationally, the generalized partial credit model is easier to handle when the estimation is based on the joint likelihood function, thanks to the exponential family form of \eqref{eq:pcm}. When the latent factors are treated as random effect, the probit version of the graded response model with multivariate normal latent factors is computationally easier to handle; see Section~\ref{subsec:mml} for more details.




\subsection{Exploratory and Confirmatory Analyses}\label{subsec:cfaefa}

Like linear factor analysis, IFA is also used in two different settings, exploratory and confirmatory settings, respectively.
Exploratory IFA assumes little or no prior knowledge about the data. It aims at learning the latent structure underlying the data through exploratory investigation procedures. Questions of interest include but not limited to: How many latent factors are needed to sufficiently describe the data? How do we interpret these factors?
The first question is a model selection problem. Given a family of IFA models, the goal is to choose the dimension $K$ of the latent factors so that the model best fits the data, in terms of model relative fit which may be measured by, for example, a suitable information criterion.

The second question is more complicated that is related to the identifiability of an exploratory IFA model. Therefore, we first discuss the model identifiability problem.
An exploratory IFA model is not identifiable for several reasons. Take model~\eqref{eq:binary} as an example, but the same reasons apply to the other models introduced above.  First, the locations and the scales of the latent factors are not identifiable. A simultaneous linear transformation of the factors and the item parameters will lead to the same model. That is, we obtain exactly the same item response function with person parameters $\tilde{\ttt}_i$ and item parameters $\tilde{\boldsymbol\beta}_j = (\tilde d_j, \tilde{\boldsymbol a}_j)$, if we let
$\tilde{\ttt}_i = \boldsymbol \mu + H \ttt_i$,  $\tilde{\mathbf a}_j = H^{-1} \mathbf a_j$, and  $\tilde{d}_j =  d_j - \mathbf a_j^\top H^{-1} \boldsymbol \mu$, for any vector $\boldsymbol\mu \in \mathbb R^{K}$ and $K\times K$ invertible diagonal matrix $H$. This location and scale indeterminacy issue is solved by imposing identification constraints. Specifically, when the latent factors are treated as fixed effects, we can fix their locations and scales by requiring
$$\sum_{i=1}^N \theta_{ik} = 0, \mbox{~and~} \frac{\sum_{i=1}^N \theta_{ik}^2}{N} = 1, k = 1, ..., K.$$
When $\ttt_i$s are treated as random effects, then these constraints are replaced by the corresponding population versions,
$$E(\theta_{ik}) = 0, \mbox{~and~} Var(\theta_{ik}) = 1, k = 1, ..., K.$$
The second unidentifiability issue of exploratory IFA is due to rotational indeterminacy. That is, even with the locations and scales of the latent factors fixed using the above identification constraints, an exploratory IFA model is still not identifiable up to an oblique rotation of the factors. \footnote{Orthogonal rotational methods (e.g., varimax rotation; \citealp{kaiser1958varimax}) are available in factor analysis that requires the estimated factors to be orthogonal to each other. As the orthogonal requirement of the latent factors is somewhat artificial, we do not discuss them in this chapter.}
More precisely, we obtain the same item response function if
we simultaneous transform the latent factors and item parameters by letting
$\tilde{\ttt}_i = H \ttt_i$, $\tilde{\mathbf a}_j =  (H^{-1})^\top \mathbf a_j$, and $\tilde d_j = d_j$,  where $H$ is a $K\times K$ invertible matrix satisfying that the diagonal entries of $HH^\top$ are all one. Note that $\tilde{\ttt}_i$ given by this transformation still follows the identification constraints, 
regardless of whether the latent factors are treated as fixed or random effects.

Dealing with the rotational indeterminacy issue requires additional assumption on the loading matrix $A$. Roughly speaking, it is often assumed that the loading matrix has a relatively simple structure, in the sense that many entries of the true loading matrix $A^*$ are zero. Specifically,  $a_{jk}^*$ being zero implies that changing the value of the $k$th factor does not change the distribution of $Y_{ij}$. When
many entries of the true loading matrix $A^*$ are zero, then each latent factor is only measured by a small set of items. If this sparsity pattern can be learned from data, then each factor can be interpreted based on the items that are directly associated with it. Different methods have been developed for learning the sparsity pattern of the loading matrix. Traditionally,  analytic rotation methods are used to find an approximate solution to this problem. An analytic rotation method starts from some estimate of the loading matrix $\hat A$, whose rotation may be fixed arbitrarily. It then finds a rotated loading matrix  $\hat AH^{-1}$ by minimizing some complexity function with respect to $H$, where the complexity function measures the sparsity of the loading matrix.
Depending on different sparsity assumptions on the true loading matrix, different complexity functions have been proposed.
We refer the readers to \cite{browne2001overview} for an overview of analytic rotation methods. It is worth noting that this type of estimators cannot produce loading matrix estimates with exactly zero entries. 
To avoid the ambiguity of analytic rotation methods for not producing exactly sparse loading estimates,
recently, penalized estimators have been developed for the learning of a sparse loading matrix \citep{sun2016latent}. These estimators obtain a sparse $\hat A$ by maximizing a penalized likelihood, where LASSO-type penalty terms are used to enforce exact sparsity.


Confirmatory IFA aims at testing substantive theory or scaling individuals along multiple latent traits, by parameter estimation, model comparison, and assessment of model goodness-of-fit. In contrast to exploratory analysis, in conducting confirmatory IFA, the number of factors, the substantive meaning of each factor, and the factors each item is measuring, are all specified a priori.
More precisely, the factors each item measures are reflected by the sparsity pattern of the loading matrix. This sparsity pattern is typically known as the design matrix or the $Q$-matrix \citep{chen2019structured} in the context of confirmatory IFA. The $Q$-matrix is a $J\times K$ binary matrix, with each entry $q_{ij} = 0$ implying that $a_{ij} = 0$ and $q_{ij} = 1$ implying that $a_{ij}$ is freely estimated.
When the zero entries in the $Q$-matrix are well positioned, then the
rotational indeterminacy issue that appears in exploratory IFA will disappear.
As a result, with further identification constraints on the location and scale of the latent factors, the loading matrix and the latent factors can be identified and consistently estimated \citep[see][for rigorous treatment on this problem]{anderson1956statistical,chen2019structured}.


\section{Estimation Methods}\label{sec:estimation}

\subsection{Estimation based on Joint Likelihood}\label{subsec:jml}

The joint maximum likelihood estimator  was first proposed in \cite{birnbaum1968some}. Treating both the latent factors and the item parameters as fixed model parameters, the joint maximum likelihood estimator is defined as
\begin{equation}\label{eq:jle}
(\hat{\ttt}_i, \hat{\boldsymbol\beta}_j: i = 1, ..., N, j = 1, ..., J) = \argmax_{\ttt_i, \boldsymbol\beta_j, i = 1, \cdots, N, j = 1, \cdots, J} \log L_{\text{J}}(\ttt_i, \boldsymbol\beta_j, i = 1, \cdots, N, j = 1, \cdots, J),
\end{equation}
where the joint likelihood function is defined in \eqref{eq:jl} and the maximization in \eqref{eq:jle} may be required to satisfy certain identification constraints though they are not explicitly stated here.

This estimator is not favored for a long period of time in the history of IFA,
which is largely due to the lack of good asymptotic properties. Under the conventional asymptotic regime that $J$ is fixed and $N$ grows to infinity, this joint estimator \eqref{eq:jle} has been shown to be inconsistent \citep{andersen1973conditional,ghosh1995inconsistent,haberman1977maximum,neyman1948consistent}.  This is not surprising, as the number of parameters grows linearly with the sample size $N$ in this asymptotic setting, which makes the classical asymptotic theory for maximum likelihood estimation fail.

However,  this traditional asymptotic regime may not be suitable for large-scale IFA applications where both $N$ and $J$ tend to be large. Instead, it may be more sensible to consider a double asymptotic setting with both $N$ and $J$ diverging. In fact, \cite{haberman1977maximum}  showed under the Rasch model that  a finite solution in \eqref{eq:jle} exists with probability tending to 1. This estimator is consistent under suitable identifiability conditions when both $N$ and $J$ diverge to infinity in suitable rates, and the estimator has asymptotic normality.
Intuitively, under this asymptotic regime, the number of parameters grows in the rate of $O(N+J)$, while the number of data points is $NJ$. The consistency is due to that the growth of the data is much faster than the dimension of the parameter space, when both $N$ and $J$ grow to infinity.

The results of \cite{haberman1977maximum} rely heavily on the convex geometry of the Rasch parameter space and thus
cannot be generalized to other IFA models. In analyzing joint-likelihood-based estimation for a wider range of IFA models, \cite{chen2019joint,chen2019structured} considered a variant of \eqref{eq:jle} called the constrained joint maximum likelihood estimator (CJMLE).
The CJMLE adds additional constraints on \eqref{eq:jle} that require $\Vert \ttt_i\Vert \leq C$ and $\Vert \boldsymbol \beta_j\Vert \leq C$, for some constant $C > 0$. These constraints restrict the estimated person parameters and item parameters to a compact ball, which not only provides technical convenience for the establishment of consistency results, but also brings numerical stability by avoiding the estimate of some person/item parameters being infinity due to the presence of extreme response patterns (e.g., $y_{ij} = 1$, for all $j = 1, ..., J$).

Due to the more complex geometry of the parameter space for general IFA models, the asymptotic results for the CJMLE are slightly weaker than that of the Rasch model. Specifically, \cite{chen2019structured} showed that for a general family of IFA models for binary data, the CJMLE leads to the convergence of the following average loss function
\begin{equation}\label{eq:loss}
\frac{\sum_{i=1}^N \sum_{j = 1}^J \left(f_{j}(1\vert \ttt_i^*, \boldsymbol\beta_j^*)- f_{j}(1\vert \hat{\ttt_i}, \hat{\boldsymbol\beta_j})\right)^2}{NJ} = O_p\left(\frac{1}{\min{\{N, J\}}}\right).
\end{equation}
As shown in \cite{chen2019structured}, when the dimension $K$ is fixed, the rate in \eqref{eq:loss} is optimal in the minimax sense.
That is, there does not exist other estimators that can beat the CJMLE in terms of the convergence rate of the loss function in \eqref{eq:loss}.
Moreover, by making use of \eqref{eq:loss} and matrix perturbation results \citep[]{davis1970rotation,wedin1972perturbation}, one can show that
under the exploratory IFA setting the loading matrix can be recovered up to a rotation, in the sense that
\begin{equation}\label{eq:loss2}
\frac{\min_{H \in \mathbb R^{K\times K}}\left\{\sum_{j=1}^J \Vert \mathbf a_j^* - H\hat{\mathbf a}_j\Vert^2\right\}}{JK} = O_p\left(\frac{1}{\min{\{N, J\}}}\right).
\end{equation}
Under the confirmatory IFA setting, when the design matrix $Q$ has a desirable structure, the convergence of the item parameters can be established as
 \begin{equation}\label{eq:loss3}
\frac{\sum_{j=1}^J \Vert \mathbf a_j^* - \hat{\mathbf a}_j\Vert^2}{JK} = O_p\left(\frac{1}{\min{\{N, J\}}}\right),
\end{equation}
which no longer involves a rotation matrix.
We remark that these results can be easily generalized to IFA models for ordinal data.


We now discuss the computation of \eqref{eq:jle} and its constrained variants. A commonly used trick to solve such a problem is by alternating maximization.  That is, one can decompose the parameters into two blocks, the person parameters and the item parameters. The optimization in
\eqref{eq:jle} or that for the CJMLE can be realized by iteratively alternating between maximizing one block of parameters given the other block. Thanks to the factorization form of the joint likelihood, the maximization in each step can be performed in parallel for different people/items. Consequently, substantial computational advantage can be gained when parallel computing techniques are used.
In addition, the maximization for each $\ttt_i$ or $\boldsymbol \beta_j$ can be viewed as finding the maximum likelihood estimate, either constrained or unconstrained, for a generalized linear regression problem. Certain IFA models have computational advantage in this step due to their exponential family forms when fixing one block of parameters.  These models include the M2PL model for binary data and the generalized partial credit model for ordinal data. Finally, we remark that the optimization problem~\eqref{eq:jle} is non-convex and thus there is no guarantee that the alternating maximization procedure converges to the global solution. Empirically, the performance of alternating maximization benefits from using a good initial value. See Section~\ref{subsec:spectral} for how a good initial value may be obtained.


\subsection{Estimation based on Marginal Likelihood}\label{subsec:mml}

In the IFA literature, it is more common to estimate the item parameters based on the marginal likelihood, where the latent factors are treated as random effects. The marginal maximum likelihood estimator (MMLE) is defined as
\begin{equation}\label{eq:mmle}
(\hat{\boldsymbol\gamma}, \hat{\boldsymbol\beta_j}, j = 1, \cdots, J) = \argmax_{\boldsymbol\gamma, \boldsymbol\beta_j, j = 1, \cdots, J}\log L_{\text{M}}(\boldsymbol\gamma, \boldsymbol\beta_j, j = 1, \cdots, J),
\end{equation}
where the marginal likelihood function is defined in \eqref{eq:ml}. In contrast to the joint-likelihood-based estimators, as the
latent factors are treated as nuisance parameters and integrated out in the likelihood function,
this estimator can be analyzed using the classical theory for maximum likelihood estimation under the traditional asymptotic regime with $J$ fixed and $N$ diverging.

Similar to the joint likelihood, the marginal likelihood function is also non-convex and thus convergence to the global solution is not guaranteed in general. The computation for \eqref{eq:mmle} tends to be much slower than \eqref{eq:jle}, due to the more complex form of the marginal likelihood.
Specifically, due to the integral in \eqref{eq:ml}, its gradient is not easy to evaluate. As a result, \eqref{eq:mmle} cannot be solved using standard numerical solvers like gradient ascent or coordinate ascent algorithms. 
The most classical method for solving \eqref{eq:mmle} is the expectation-maximization (EM) algorithm \citep{bock1981marginal,dempster1977maximum}. For ease of explanation, we simplify the notation by denoting $\Psi = (\boldsymbol\gamma, \boldsymbol\beta_j, j = 1, \cdots, J)$. Starting from an initial set of  parameters $\Psi^{(0)}$, the EM algorithm iteratively performs two steps, the Expectation (E) step and the Maximization (M) step.
In the $t+1$th iteration, the E step constructs an objective function $Q(\Psi\mid \Psi^{(t)})$ given the parameter values from the previous step $\Psi^{(t)}$, where
\begin{equation}\label{eq:em}
Q(\Psi\mid \Psi^{(t)}) = \sum_{i=1}^N E_{\ttt_i\mid \Psi^{(t)},\mathbf{y}_i}\left[ \log\phi(\ttt_i\vert \boldsymbol \gamma)+\sum_{j=1}^J \log f_j(y_{ij}\vert \ttt_i; \boldsymbol\beta_j)\right].
\end{equation}
The expectation in \eqref{eq:em} is with respect to the latent factors $\ttt_i$ under its posterior distribution based on the current parameters $\Psi^{(t)}$. Then the M step produces the parameter values $\Psi^{(t+1)}$ by
maximizing $Q(\Psi\mid \Psi^{(t)})$, i.e.,
$$\Psi^{(t+1)} = \argmax_{\Psi}Q(\Psi\mid \Psi^{(t)}).$$
With a good starting point $\Psi^{(0)}$, the sequence $\Psi^{(t)}$ will converge to the marginal maximum likelihood estimate as defined in \eqref{eq:mmle}.

When the dimension $K$ of the latent factors is large, the computational complexity of the E step is high, for
the above vanilla version of the EM algorithm. This is because, the expectation with respect to $\ttt_i$ involves a $K$-dimensional integral that does not have a closed form. As a result, the expectation has to be evaluated by a numerical integral, whose complexity is exponential in the dimension $K$. Even for a moderate $K$ (e.g., $K > 5$), the computation of the vanilla EM algorithm can hardly be affordable.

In dealing with the computation of large-scale IFA under the random effect view of the latent factors, several methods have been proposed. One way to improve the EM algorithm is by replacing the numerical integral in the E step using Monte Carlo integration, which leads to the Monte Carlo EM algorithm \citep{meng1996fitting,wei1990monte} for IFA. The Monte Carlo EM algorithm approximates the $Q$ function by approximating the expectations in \eqref{eq:em} by Monte Carlo integration, which means to sample from the posterior distribution of each $\ttt_i$ under the current model $\Psi^{(t)}$. As the posterior distribution of $\ttt_i$ typically does not have a simple closed form, Markov chain Monte Carlo (MCMC) methods are often used.
Even with the numerical integration avoided,
the Monte Carlo EM still tends to be slow in practice. In a later stage of the Monte Carlo EM algorithm, the error in $\Psi^{(t)}$ will be dominated by the Monte Carlo error from the E step. To accurately approximate the MMLE,  
a very large number of Monte Carlo samples is needed in the last several iterations of the algorithm, which can cause a very high computational burden.

The computational inefficiency of the Monte Carlo EM algorithm is largely due to the inefficient use of the posterior samples of the latent factors. That is, the posterior samples of the latent factors are only used once and are immediately discarded in the next iteration.
To avoid the inefficient use of the posterior samples, alternative methods have been proposed, including the stochastic EM algorithm \citep{celeux1985sem,ip2002single,nielsen2000stochastic,zhang2018improved} and stochastic approximation with MCMC algorithms \citep{cai2010high,cai2010metropolis,gu1998stochastic}. The stochastic EM algorithm differs from the EM algorithms by the following ways. Let $\Psi^{(t)}$ be the parameter estimate from the $t$th iteration. 
In the $(t+1)$th iteration, the stochastic EM algorithm replaces the E step of the EM algorithm by a stochastic E step.
In this stochastic E step, we
obtain one sample $\ttt_i^{(t+1)}$ for each $\ttt_i$ from its posterior distribution under the current model, then construct
a $Q$-function in the form
$$Q(\Psi\mid \Psi^{(t)}) = \sum_{i=1}^N  \left[\log\phi(\ttt_i^{(t+1)}\vert \boldsymbol \gamma)+\sum_{j=1}^J \log f_j(y_{ij}\vert \ttt_i^{(t+1)}; \boldsymbol\beta_j)\right].$$
Then in the M step of stochastic EM algorithm, we obtain
$$\Psi^{(t+1)} = \argmax_{\Psi}Q(\Psi\mid \Psi^{(t)}).$$
Unlike the EM and Monte Carlo EM algorithms for which the final estimate is given by $\Psi^{(T)}$ from the last iteration,
the final estimate of the stochastic EM algorithm is given by an average of $\Psi^{(t)}$s from the iterations after a sufficient burn-in period, i.e.,
\begin{equation}\label{eq:stEM}
\hat \Psi = \frac{\sum_{t=m+1}^T \Psi^{(t)}}{T-m},
\end{equation}
where iterations 1 to $m$ are used as a burn-in period and the last $T-m$ iterations are used in the final estimate $\hat \Psi$. As shown in \cite{nielsen2000stochastic}, the estimator \eqref{eq:stEM} is almost asymptotically equivalent to the MMLE for a sufficiently large number of iterations $T$. Comparing with the Monte Carlo EM algorithm, the stochastic EM algorithm more efficiently makes use of the posterior samples of the latent factors, by including  $\Psi^{(t)}$ from many iterations in the final estimate.
The theoretical properties of the stochastic EM algorithm are
studied comprehensively in \cite{nielsen2000stochastic}. Computational details and numerical examples of applying stochastic EM to large-scale IFA problems can be found in \cite{zhang2018improved}.

The stochastic approximation with MCMC algorithms \citep{cai2010high,cai2010metropolis,gu1998stochastic} are developed based on the seminal work of \cite{robbins1951stochastic} on stochastic approximation. It is worth pointing out that the work of \cite{robbins1951stochastic} also lays the theoretical foundation for the gradient descent optimization method that is widely used in machine learning and artificial intelligence.
Similar to the EM algorithm and its variants as introduced above, a stochastic approximation MCMC algorithm also iteratively alternates between two steps. Let the current iteration number be $t$ and the current parameter estimate be $\Psi^{(t)}$. To proceed, the algorithm first does a stochastic E step. In this step, a small number of posterior samples for each $\ttt_i$ are obtained under the current model $\Psi^{(t)}$. These posterior samples are used to find a stochastic gradient of the marginal likelihood \eqref{eq:ml} at the current parameter $\Psi^{(t)}$.
We denote this stochastic gradient as $H^{(t+1)}$. In some algorithms, an approximation to the Hessian matrix of \eqref{eq:ml} is also obtained in this step. We denote the Hessian matrix approximation by $\Gamma^{(t+1)}$. Then following the idea of stochastic approximation, in the second step of the iteration, the parameters are updated by a stochastic gradient ascent step
\begin{equation}\label{eq:stoappo}
\Psi^{(t+1)} = \Psi^{(t)} + \gamma_{t+1} H^{(t+1)},
\end{equation}
or
\begin{equation}\label{eq:stoappo2}
\Psi^{(t+1)} = \Psi^{(t)} + \gamma_{t+1} \left[\Gamma^{(t+1)}\right]^{-1} H^{(t+1)},
\end{equation}
where $\gamma_t$s is a pre-specified constant that is typically chosen as  $\gamma_t = 1/t$. This constant is known as the gaining constant in the stochastic approximation literature and it plays a very important role in the algorithm. Comparing the two
updating methods \eqref{eq:stoappo} and \eqref{eq:stoappo2}, the second updating rule may lead to faster convergence when the Hessian matrix is accurately approximated \citep{chung1954stochastic,fabian1968asymptotic}. However, it can also lead to numerical instability and does not improve the convergence when the approximation is poor. 
This type of algorithms was first proposed by \cite{gu1998stochastic} for handling general missing data problems. It is then tailored to
large-scale IFA problems in \cite{cai2010high,cai2010metropolis}, with computational details and numerical examples provided.


A key step, which is also the most time-consuming part of all the above variants of the EM algorithm, is to sample from the posterior distribution of $\ttt_i$ under the current set of parameters. This problem can be non-trivial due to the lack of conjugacy, in particular, when the dimension $K$ of the latent factors is high. Although the Gibbs sampler can always be used, it is likely to suffer from the slow-mixing issue, especially when $K$ is large and the factors are highly correlated. However, there is one family of models, for which the posterior samples of $\ttt_i$ are easy to obtain. These models include model \eqref{eq:binary} and the graded response model \eqref{eq:ordinal}, when the probit link is used and the marginal distribution $F$ is multivariate normal. Under these models, $\ttt_i$ can be sampled using a blocked Gibbs sampler.
This blocked Gibbs sampler is designed using a data argumentation trick that is based on the latent response formulation. Take the model for binary data as an example. We consider the sampling of
$\ttt_i$ given data $y_{i1}$, ..., $y_{iJ}$ and model parameters $\boldsymbol\gamma$, $\boldsymbol\beta_{1}$, ..., $\boldsymbol\beta_{J}$.
The Gibbs sampler iterates between the following two steps, both of which can be easily implemented.
\begin{itemize}
  \item[] Step 1: Independently sample latent responses $y_{ij}^*$, $j = 1, ..., J$. Each $y_{ij}^*$ is sampled from a unidimensional truncated normal distribution with density function
      $$g(y)\propto \left\{\begin{array}{ll}
              \exp\left(-(y - d_j - \mathbf a_j^\top \ttt_i)^2/2\right) 1_{\{y\geq 0\}},& \mbox{~if~} y_{ij} = 1, \\
              \exp\left(-(y - d_j - \mathbf a_j^\top \ttt_i)^2/2\right)1_{\{y< 0\}}, & \mbox{~if~} y_{ij}  = 0,
            \end{array}\right.$$
            where the value of $\ttt_i$ is from the previous step.
  \item[] Step 2: Update $\ttt_i$ by sampling from a multivariate normal distribution
  $$h(\ttt) \propto \phi(\ttt\vert \boldsymbol\gamma)\prod_{j=1}^J \exp\left(-(y_{ij}^* - d_j - \mathbf a_j^\top \ttt)^2/2\right),$$
  where $y_{ij}^*$s are from Step 1.
\end{itemize}

We provide a brief comparison of the three stochastic variants of the EM algorithm. In terms of computational speed,
the stochastic approximation with MCMC algorithm and the stochastic EM algorithm are similar in achieving the same accuracy, and the former may be slightly faster in some situations. Both methods tend to be substantially faster than the Monte Carlo EM algorithm. Further comparing the stochastic approximation method and the stochastic EM algorithm, the latter tends to be numerically more stable and thus easier to use for applied researchers.
This is because, the   stochastic EM algorithm is almost tuning free. In contrast, the performance of the stochastic approximation method is sensitive to the specification of the gaining constant and the accurate approximation of the Hessian matrix (when using updating rule \eqref{eq:stoappo2}). As a result, good performance of the stochastic approximation method is sometimes subject to tuning, which can be labor intensive.



Besides the variants of the EM algorithms described above, full Bayesian methods have also been developed to find approximate solutions to \eqref{eq:mmle}. A full Bayesian method imposes prior distributions on the parameters in \eqref{eq:ml}, including $\boldsymbol \gamma$ and $\boldsymbol\beta_j$, $j = 1, ..., J$. Then posterior means/modes of these parameters are approximated using Markov Chain Monte Carlo Methods.
See \cite{beguin2001mcmc}, \cite{bolt2003estimation}, and \cite{edwards2010markov} for applications of full Bayesian methods under various IFA settings. We point out that most applications of IFA take a frequentist setting. Even when full Bayesian methods are used for the computation, they tend to be used as a tool to approximate the marginal maximum likelihood estimator, making use of the asymptotic equivalence  between posterior mean/mode and the maximum likelihood estimator in the frequentist sense \citep[see e.g., Chapter 10,][]{van2000asymptotic}.

\subsection{Limited-information Estimation}

Methods introduced in Sections~\ref{subsec:jml} and \ref{subsec:mml} are typically known as the full-information estimation methods, as both the joint likelihood function and marginal likelihood function are based on the joint distribution of data. In contrast, limited-information estimation methods only make use of limited information such as univariate and bivariate proportions.  In what follows, we review two commonly used limited-information methods.

The first method is called the composite-likelihood-based estimator \citep[]{joreskog2001factor,zhao2005composite}. This estimator is based on the marginal distributions of univariate and bivariate responses, in which the latent factors are still treated as random effects.
This approach applies to all the models introduced above, as long as the latent factors  follow a multivariate normal distribution.
More precisely, the estimator is obtained by maximizing the following composite likelihood function
\begin{equation}\label{eq:comp1}
(\hat{\boldsymbol\gamma}, \hat{\boldsymbol\beta_j}, j = 1, \cdots, J) = \argmax_{\boldsymbol\gamma, \boldsymbol\beta_j, j = 1, \cdots, J}\log L_{\text{C}}(\boldsymbol\gamma, \boldsymbol\beta_j, j = 1, \cdots, J),
\end{equation}
where
\begin{equation}\label{eq:comp2}
\begin{aligned}
L_{\text{C}}(\boldsymbol\gamma, \boldsymbol\beta_j, j = 1, \cdots, J) = &\prod_{i=1}^N\Bigg\{ \left[\prod_{j=1}^{J-1} \prod_{j' = j+1}^J \int f_j(y_{ij}\vert \ttt; \boldsymbol\beta_j) f_{j'}(y_{ij'}\vert \ttt; \boldsymbol\beta_{j'})\phi(\ttt\vert \boldsymbol\gamma)d\ttt\right]\\
&\times \left[\prod_{j=1}^{J}\int f_j(y_{ij}\vert \ttt; \boldsymbol\beta_j)\phi(\ttt\vert \boldsymbol\gamma)d\ttt\right]\Bigg\}.
\end{aligned}
\end{equation}
This is a composite likelihood function as a product of all the univariate and bivariate likelihoods.
When the latent factors follow a multivariate normal distribution,
the computation of the estimator \eqref{eq:comp1} tends to be much easier than that of \eqref{eq:mmle} when the dimension $K$ is large. This is because the integrals in \eqref{eq:comp2} can be simplified to one- or two-dimensional integrals by a change of variables.
For example,
the integral
$$\int f_j(y_{ij}\vert \ttt; \boldsymbol\beta_j) f_{j'}(y_{ij'}\vert \ttt; \boldsymbol\beta_{j'})\phi(\ttt\vert \boldsymbol\gamma)d\ttt$$
is two-dimensional, if we integrate with respect to $(\mathbf a_j^\top\ttt, \mathbf a_{j'}^\top\ttt)^\top$, which follows a bivariate normal distribution.


When the number of items $J$ is large, the computation of \eqref{eq:comp1} tends to be slower than the CJMLE, stochastic EM and stochastic approximation with MCMC algorithms. The gap in computational speed becomes even larger when parallel computing is allowed, as efficient parallel computing algorithms can be designed for the other methods but not the composite-likelihood-based estimator. This is because the total number of parameters is in the order of $J$. All these parameters have to be optimized together in \eqref{eq:comp1} due to the form of the objective function, while the optimizations and updates in the other algorithms can be decomposed into many small problems that can be performed in parallel.
For example, in the alternating maximization of the joint likelihood, the maximization in each step can be performed independently for different people/items.


Under the classical asymptotic regime, the consistency and asymptotic normality of this estimator hold under standard assumptions, following the general theory for composite likelihood estimation \citep[see][for an overview]{varin2011overview}. Of course, comparing with the MMLE based on the full marginal likelihood, the composite-likelihood-based estimator tends to be statistically less efficient, for only using univariate and bivariate information.

The second method is based on the concept of polychoric/tetrachoric correlation for binary/ordinal data \citep{joreskog1994estimation,lee1990three,muthen1984general}. This method makes use of the connection between the linear factor model and IFA models and only applies to model \eqref{eq:binary} and the graded response model \eqref{eq:ordinal} with a probit link and multivariate normal $F$. These models have a latent response interpretation and the latent responses $Y_{i1}^*$, ..., $Y_{iJ}^*$ are multivariate normal. The correlation between each pair $(Y_{ij}^*, Y_{ij'}^*)$ is known as the
tetrachoric correlation when both variables are binary and the polychoric correlation when one or both of them are ordinal.
This correlation can be consistently estimated based on observed data from the pair, $(y_{ij}, y_{ij'}), i = 1, ..., N$.
By the definition of latent response (e.g., equation \eqref{eq:uv} for the binary case), the covariance matrix of the latent responses takes the form $A\Phi A^\top + I$, which further implies the polychoric/tetrachoric correlation matrix. Here,  $\Phi$ is the covariance of $\ttt_i$ and $I$ is a $K\times K$ identity matrix. With suitable identification constraints, the loading matrix $A$ can be consistently estimated by minimizing a certain distance between the estimated polychoric/tetrachoric correlation matrix and the model implied one. Like the composite-likelihood-based estimator above, under the classical asymptotic regime, this method also tends to be statistically less efficient than the MMLE.


\subsection{Spectral Method}\label{subsec:spectral}

In linear factor analysis, principal component analysis (PCA) is commonly used as a spectral method for exploratory analysis.
This approach is computationally much faster and does not suffer from convergence issue as in many other methods for involving nonconvex optimization, as it only involves eigenvalue decomposition. Moreover, thanks to a close connection between PCA and linear factor models, the PCA solution to exploratory factor analysis is statistically consistent when data follow a linear factor model and both the sample size and the number of variables grow to infinity \citep{stock2002forecasting}. Thanks to both the computational advantage and theoretical guarantee, PCA is often the first estimation method to apply when conducting exploratory linear factor analysis.
The PCA solution is also often used as the starting point in the computation of other estimators, which are often used after the PCA for providing more accurate estimation results and uncertainty quantification.

PCA cannot be used in IFA. Fortunately, a singular value decomposition (SVD) method has been proposed that can play a
similar role as PCA in linear factor analysis \citep{zhang2019note}.
Similar to PCA, the SVD method only involves singular value decomposition. Thus, it is computationally faster than other estimation methods and does not suffer from convergence issues. This method is built based on the SVD method for matrix estimation first proposed in \cite{chatterjee2015matrix}. In what follows, we discuss the main steps of this SVD method for analyzing binary data.
\begin{itemize}
  \item[]Step 1: Apply singular value decomposition to the binary data matrix $(y_{ij})_{N\times J}$.
  \item[]Step 2: Denoise by truncating the small singular values to zero and obtain $\hat P = (\hat p_{ij})_{N\times J}$ as an estimate of the response probability matrix $(f_{j}(1\vert \ttt_i, \boldsymbol\beta_j))_{N\times J}$.
  \item[]Step 3: Obtain an estimate $\bar M$ of the matrix $(\mathbf a_j^\top \ttt_i)_{N\times J}$ by truncating and transforming $\hat P$.
  \item[]Step 4: Obtain estimates $\hat \ttt_i$, $\hat{\mathbf a}_j$, and $d_j$, $i = 1, ..., N, j = 1, ..., J$  by singular value decomposition of $\bar M$.
\end{itemize}
This method can also be used to analyze ordinal data, by dichotomizing the ordinal data to multiple binary data matrices.


As shown in \cite{zhang2019note}, this method is also statistically consistent, in the sense that ${\min_{H \in \mathbb R^{K\times K}}\left\{\sum_{j=1}^J \Vert \mathbf a_j^* - H\hat{\mathbf a}_j\Vert^2\right\}}/{(JK)} = o_p(1)$,
when both $N$ and $J$ grow to infinity, under similar conditions as those needed for the consistency of CJMLE \citep{chen2019joint,chen2019structured}. 
However, it is worth pointing out that the SVD method sacrifices statistical efficiency. Under suitable conditions, it can be shown that the loading matrix estimate $\hat A$ achieves the convergence rate
\begin{equation}\label{eq:loss4}
\frac{\min_{H \in \mathbb R^{K\times K}}\left\{\sum_{j=1}^J \Vert \mathbf a_j^* - H\hat{\mathbf a}_j\Vert^2\right\}}{JK} = O_p\left(\frac{1}{(\min{\{N, J\}})^{\frac{1}{K+2}}}\right),
\end{equation}
while that same loss has the rate $O_p(1/\min{\{N, J\}})$ for the CJMLE. Thus, this SVD method is more suitable as an initial step for exploratory IFA to provide researchers a first impression on the data structure and to provide a starting point for the computation of other estimators.

\section{Computer Implementations} 
\label{sec:computation_imple}

In this section, we briefly introduce the commonly used statistical softwares and packages for IFA estimation.
In the last decade, the large-scale IFA problem has received much attention in statistics and psychometrics and many computer implementations of IFA methods have emerged. In particular, several R packages have been developed and well-maintained that provide researchers computationally efficient tools for solving large-scale IFA problems. However, one has also to admit that
the existing statistical softwares/packages have their own focus either on estimation methods or IFA models.
There still lacks a statistical software/package that can implement all the state-of-the-arts methods for all commonly used IFA models.


\paragraph{Commercial Softwares} 
\label{par:commercial_softwares}
\begin{itemize}
    \item \textbf{Mplus} \citep[Version 8.4;][]{muthen2020mplus} is a general latent variable modeling program available on major platforms including Windows, MacOS, and Linux. In particular, it is capable of doing exploratory and confirmatory factor analysis for binary or ordinal responses using weighted least square method, Gauss-Hermite EM algorithm, or full Bayesian approach via MCMC. A comparison of Mplus and IRTPRO for high dimensional item factor analysis can be found in \cite{asparouhov2012comparison}.
    \item \textbf{PARSCALE} \citep[Version 4.10;][]{parscale} is a Windows platform software for unidimensional item factor analysis. It is applicable for both binary and ordinal responses.
    \item \textbf{BILOG-MG} \citep[Version 3.00;][]{bilogmg} is a Windows platform software for binary response IRT analysis. It is an extension of BILOG with multiple-group respondents support.
    \item \textbf{IRTPRO} \citep[Version 4.20;][]{cai2011irtpro} is a Windows-based software. It can be used for unidimensional and multidimentional item response theory models like 2PL, 3PL, generalized partial credit (GPC) model in both exploratory and confirmatory analysis. Multiple algorithms are implemented in IRTPRO such as Gauss-Hermite quadrature EM \citep[]{bock1981marginal}, MH-RM \citep[]{cai2010metropolis}, MCMC \citep[]{patz1999applications}, etc.
    \item \textbf{flexMIRT} \citep[Version 3.5;][]{cai2013flexmirt} is a Windows-based IRT analysis software. It works for unidimensional and multidimentional IRT models including 1-3PL models for dichonomous response, graded response model, generalized partial credit model for polytomous response through MML estimation. Gauss-Hermite EM, MH-RM algorithms are used in flexMIRT for the optimization procedure.
\end{itemize}
\paragraph{Free softwares}
\begin{itemize}
  \item \textbf{WinBUGS} \citep[Version 1.4;][]{spiegelhalter2003winbugs} is a free software aiming at the Bayesian analysis on the Windows platform. It belongs to the BUGS (Bayesian inference Using Gibbs Sampling) project together with OpenBUGS. WinBUGS can be used for full Bayesian item factor analysis.
\end{itemize}
\paragraph{R Packages (open source)} 
\label{par:r_packages_}
\begin{itemize}
    \item \textbf{ltm} \citep[Version 1.1-1;][]{rizopoulos2006ltm} is developed for the MML estimation of IRT models with Rasch, 2PL, 3PL, and Graded Response models, using Gauss-Hermite EM algorithm. It is written in pure R and it uses BFGS algorithm implemented `optim()' function in R's base \textbf{stats} package for the optimization.
    \item \textbf{eRm} \citep[Version 1.0-0;][]{mair2007extended} is written in mixed C, Fortran, and R and focuses on the estimation of extended Rasch models using conditional maximum likelihood (CML) method.
    \item \textbf{TAM} \citep[Version 3.3-10;][]{TAM_3.3-10} is written in mixed C++, R. It could perform item response modeling under a variaty of models, including Rasch model, 2PL/3PL model, generalized partial credit model, etc. It uses quasi Monte Carlo (QMC) integration to prevent computational demanding in ordinary Gaussian quadrature integration.
    \item \textbf{MCMCpack} \citep[Version 1.4-4;][]{martin2011mcmcpack} is written in C++, R and aims for Bayesian inference by drawing MCMC samples from posterior under eighteen statistical models including unidimensional and multidimentional IRT models. It is similar to BUGS and JAGS system but uses compiled and tailored code for the estimation of several pre-specified models, through which efficiency is gained.
    \item \textbf{mirt} \citep[Version 1.31;][]{chalmers2012mirt} is written in C++ and R. It contains functions for MML estimation of IRT models including Rasch, 2PL, and 3PL, ordinal response models, in an exploratory or confirmatory analysis. Gauss-Hermite quadrature EM, MH-RM, and stochastic EM algorithms are implemented in the package.
    \item \textbf{lvmcomp} \citep[Version 1.2;][]{lvmcomp} is written in C++ and R. It implementes the improved stochastic EM algorithm for full-information item factor analysis \citep[]{zhang2018improved}. Boosted by parallel computing technique through the C++ OpenMP API \citep[]{dagum1998openmp}, the \textbf{lvmcomp} package is especially suitable for medium-to-large scale estimation problems for multidimentional 2-parameter logistic (M2PL) and generalized partial credit (GPC) model under the confirmatory setting.
    \item \textbf{mirtjml} \citep[Version 1.3.0;][]{mirtjml} is written in C++ and R. Through the efficient constrainted joint maximum likelihood estimation \citep[CJMLE,][]{chen2019joint,chen2019structured} algorithm and parallel computing technique, the \textbf{mirtjml} package is powerful in item factor analysis when the sample size, the number of items, and the number of factors are all large. For now, it provides functions for exploratory and confirmatory item factor analysis under the M2PL model.
\end{itemize}

\section{Conclusions}

In this chapter, we reviewed the modeling framework for IFA and the state-of-the-arts methods for its estimation. Our focus was on large-scale IFA applications, where the sample size, the number of items, and the number of factors are all large.
In the recent decades, many works have been done on this topic and we believe that the problem of efficiently obtaining a point estimate has been solved well. Our tool box is now equipped with several powerful tools. For very large-scale problems, we tend to use
the SVD method and the joint maximum likelihood estimation for a fast investigation of the data. For moderate-size problems, we would suggest to treat the person parameters as random effects and estimate the loading parameters by marginal-likelihood-based methods, such as MCMC, stochastic EM, and stochastic approximation with MCMC. Among these marginal-likelihood-based methods, the stochastic EM tends to achieve a better balance between computational efficiency and numerical stability. Several computer softwares/packages are available and well-maintained, though  a comprehensive computation platform is needed  to better support applied research  that aggregates all these estimation methods under a wide range of models.


What has not been well-solved is the uncertain quantification for estimated IFA models (e.g., constructing confidence intervals/regions), especially for large-scale problems. The challenge lies in that when the sample size, the number of items, and the number of factors are all large, the classical asymptotic normality theory may no longer apply. New inference methods and asymptotic theory remain to be developed under a new regime that the sample size, the number of items, and possibly also the number of factors diverge.
This is a challenging problem in general.
Ideas from high-dimensional statistical inference \citep[e.g.,][]{chernozhukov2018double} may be borrowed to solve the problem.

\bibliographystyle{apa}
\bibliography{ref}

\end{document}